\documentclass[preprint]{aastex63}
\usepackage{graphicx}
\accepted{for publication in the Astronomical Journal}

\begin{document}
\title{KK 242, a faint companion to the isolated Scd galaxy NGC 6503}
\author{Igor D. Karachentsev}
\affil{Special Astrophysical Observatory, the Russian Academy of Sciences, Nizhnij Arkhyz, Karachai-Cherkessian Republic,
   Russia 369167}
\email{ikar@sao.ru}
   
\author{John M. Cannon}
\affil{Department of Physics \& Astronomy, Macalester College,
  1600 Grand Avenue, Saint Paul, MN 55105, USA}

\author{Jackson Fuson}
\affil{Department of Physics \& Astronomy, Macalester College,
  1600 Grand Avenue, Saint Paul, MN 55105, USA}

\author{John L. Inoue}
\affil{Department of Physics \& Astronomy, Macalester College,
  1600 Grand Avenue, Saint Paul, MN 55105, USA}

\author{R. Brent Tully}
\affil{Institute for Astronomy, University of Hawaii, 2680 Woodlawn Drive, 
Honolulu, HI 96822, USA} 

\author{Gagandeep S. Anand}
\affil{Space Telescope Science Institute, 3700 San Martin Drive, Baltimore, MD 21218, USA}
    
\author{Serafim S. Kaisin}
\affil{Special Astrophysical Observatory, the Russian Academy of Sciences, Nizhnij Arkhyz, Karachai-Cherkessian Republic,
  Russia 369167}

     
\begin{abstract}

Using Hubble Space Telescope imaging of the resolved stellar
population of KK~242 = NGC6503-d1 = PGC~4689184, we measure the
distance to the galaxy to be $6.46\pm0.32$ Mpc and find that KK~242 is
a satellite of the low-mass spiral galaxy NGC~6503 located on the edge
of the Local Void. Observations with the Karl G. Jansky Very Large
Array show signs of a very faint HI-signal at the position of KK~242
within a velocity range of $V_{hel} = -80\pm10$ km\,s$^{-1}$.  This
velocity range is severely contaminated by HI emission from the Milky
Way and from NGC6503. The dwarf galaxy is classified as the transition
type, dIrr/dSph, with a total HI-mass of $< 10^6 M_{\odot}$ and a
star formation rate SFR(H$\alpha$) = --4.82 dex ($M{\odot}$/yr). Being
at a projected separation of 31 kpc with a radial velocity difference
of -- 105 km\,s$^{-1}$ relative to NGC~6503, KK~242 gives an estimate
of the halo mass of the spiral galaxy to be $\log(M/M_{\odot}$) =
11.6. Besides NGC~6503, there are 8 more detached low-luminosity
spiral galaxies in the Local Volume: M~33, NGC~2403, NGC~7793,
NGC~1313, NGC~4236, NGC~5068, NGC~4656 and NGC~7640, from whose small
satellites we have estimated the average total mass of the host
galaxies and their average total mass-to-K-band-luminosity $\langle
M_T/M_{\odot}\rangle = (3.46\pm0.84)\times 10^{11}$ and $(58\pm19)
M_{\odot}/L_{\odot}$, respectively.

\end{abstract}
  

\section{Introduction.}

The ratio of the stellar mass of a galaxy, $M_*$, to its halo mass, $M_h$,
is a function of the integral luminosity and morphological type of the
galaxy. The minimum $M_h/M_*$ value is characteristic of spiral galaxies
like the Milky Way and M~31, and the value $M_h/M_*$ is  
increasing towards objects of both high and low
luminosity (Kourkchi \& Tully, 2017). The presence of such a minimum is 
usually explained by the high efficiency of star formation exactly in galaxies
such as Milky Way and M~31 (Correa \& Schaye 2020). The $M_h/M_*$ ratio for
early-type galaxies (E, S0) is 2--3 times higher than that for spirals of the
same stellar mass (Karachentseva et al. 2011, Posti \& Fall 2021, Bilicki 
et al. 2021). This is probably due to the fact that in the last 10 Gyr,
star formation in E, S0 galaxies took place at a very low rate.

Unlike massive spiral galaxies, the characteristic $M_h/M_*$ value for low-
luminosity spiral galaxies has not yet been reliably determined. According to
Lapi et al, (2018), the disk-like galaxies with luminosities an order of
magnitude lower than the Milky Way luminosity and rotation amplitudes
$V_{rot}<120$~km s$^{-1}$ have mainly increasing rotation curves. Extrapolation of 
the value $V_{rot}$, measured within the optical radius of the galaxy ($\sim20$~ kpc),
to a distance equal to the virial radius of the halo ($\sim120$~kpc) introduces a
significant uncertainty in the estimate of $M_h$.

The halo mass of a galaxy can be determined from the relative radial
velocities and projected separations of its satellites. However, the
number of satellites around low-mass galaxies is small. On average,
there are only 1--2 dwarf satellites per one late-type spiral galaxy
(Sc--Sd) with a luminosity in the K-band of
$L_K/L_{\odot}\simeq(9.5-10.0)$~dex. For example, the M 33 galaxy has
2 probable satellites: And XXII (Martin et al. 2009, Tollerud et
al. 2012) and Tri III (Martinez Delgado et al. 2021).  Carlin et
al. (2016) undertook a search for faint satellites around another
nearby isolated spiral galaxy of moderate luminosity, NGC 2403. To its
collection of known satellites (DDO 44 and NGC 2366) they added one
new object, MADCASHJ0742+65, the radial velocity of which has not yet
been measured.  Carlsten et al. (2020) has performed deep searches for
new companions around 10 nearby massive galaxies. However, this study
was limited to virial neighborhoods of galaxies with luminosities
comparable to those of Milky Way and M 31. Moderate luminosity
galaxies were not considered.

Karachentseva \& Karachentsev (1998) used the photografic Palomar sky
survey POSS-II to search for nearby dwarf galaxies over the whole
sky. Near the Scd galaxy NGC~6503 at a separation of $17\arcmin$, they
found an object of low surface brightness, called KK~242. Huchtmeier
et al. (2000) observed it in the 21 cm HI-line with the 100-meter
Effelsberg telescope and detected its HI-flux, 2.03 Jy,~km s$^{-1}$ at
the radial velocity of $V_h = 426$~km s$^{-1}$ with the HI-line width
of $W_{50}= 100$~km s$^{-1}$ km/s. The obtained radial velocity
exceeds the radial velocity of NGC~6503, $V_h= +25$~km s$^{-1}$
(Epinat et al. 2008, Greisen et al. 2009), over $\sim400$~km s$^{-1}$,
and the high HI-flux and the wide HI-line width do not harmonize with
the low luminosity of KK~242 at its apparent magnitude $B = 18\fm6$.
Later, Koda et al. (2015) performed a deep survey of the vicinity of
NGC~6503 on the Subaru telescope with Suprime-Camera and rediscovered
the KK~242 dwarf system, giving it the name NGC6503-d1. On the deep
images in B, V, R, I-bands the dwarf galaxy KK~242 = NGC6503-d1 = PGC
4689184 was resolved into stars. Based on the color-magnitude diagram for
them, the authors estimated the dwarf galaxy distance to be at a
distance of $\sim$5.3 Mpc. With an error in estimating the distance of
$\sim 1$ Mpc, this value is consistent with the distance of the galaxy
NGC~6503 itself, $D = 6.25$ Mpc (Extragalactic Distance Database =
EDD, http://edd.ifa.hawaii.edu), measured from V, I- images on the
Hubble Space Telescope (HST). However, to confirm the physical
connection between KK~242 and NGC~6503, a more accurate measurement of
the distance to KK~242 is necessary, as well as verification and
refinement of its radial velocity. The measurement of these parameters
is the subject of our work.\clearpage

 \section{The TRGB distance derived with HST.}  

The first observations of KK~242 with the Hubble Space Telescope 
were made on October 4, 1999 under the SNAP program 8192 (PI P.Seitzer).
Two images of the galaxy were obtained in F606W (600 s) and F814W (600 s)
filters with WFPC2 camera. The galaxy was resolved into stars, but their
color-magnitude diagram (CMD) turned out to be too shallow to determine
the galaxy distance.

New observations of KK~242 were performed with the Advanced
Camera for Surveys (ACS) aboard the HST on October 15, 2019 as 
part of the ''Every Known Nearby Galaxy'' survey (SNAP-15922, PI 
R.B. Tully). Two exposures were made in a single orbit 
with the filters F606W (760 s) and F814W (760 s). An inverted 
color cutout of the galaxy produced with this data is 
shown in Figure 1, with a size of 104$\arcsec$ by 70$\arcsec$.
The galaxy contains young and older stellar population visible as
red and blue dots, respectively.

We used the ACS module of the DOLPHOT package
\footnote{http://purcell.as.arizona.edu/dolphot} by Dolphin (2000,
2016) to perform photometry of the resolved stars based on the
recommended recipe and parameters. Only stars with good-quality
photometry (defined as type $\leq 2$) were included in the analysis.
We also selected stars with a signal-to-noise ratio of at least five
in both filters, and with DOLPHOT parameters (Crowd$_{\rm F606W}$ +
Crowd$_{\rm F814W}) \leq0.8$), (Sharp$_{\rm F606W}$ + Sharp$_{\rm
  F814W})^2 \leq0.075$ (McQuinn 2017).  As the galaxy does not take up
the entire 202$\arcsec$ by 202$\arcsec$ ACS field of view, we isolated
the galaxy to produce a CMD which is limited in contamination from
objects such as foreground stars or unresolved background galaxies.

Artificial stars were inserted and recovered using the same DOLPHOT
parameters to accurately estimate the photometric errors.  The
resulting colour-magnitude diagram in F606W--F814W versus F814W is
plotted in Figure 2, shown along with representative error bars from
the results of the artificial star experiments.  We measured the
magnitude of the tip of the red giant branch (TRGB) by following the
methods outlined by Makarov et al. (2006) and Wu et al. (2014). We
modeled the luminosity function of asymptomtic giant branch and red
giant branch stars as a broken power-law, with the break defining the
magnitude of the TRGB. The physical reason for this parameterization
is that the abrupt and standard-candle nature of the onset of the
helium flash presents itself as a discontinuity on the observed
luminosity function.  The benefit of this procedure over a simple
edge-detection algorithm (e.g. a Sobel filter) is the ability to
directly incorporate the results of artificial star experiments. This
is important as DOLPHOT has been shown to systematically underestimate
the reported photometric errors (e.g. Williams et al., 2014). The
results from the artificial star experiments are explicitly taken into
account by convolving the pre-defined luminosity function with
functions that account for completeness, photometric error, and
bias. Using this procedure, we found F814W(TRGB) to be
$25\fm01\pm0\fm11$.  Following the zero-point calibration of the
absolute magnitude of the TRGB developed by Rizzi et al. (2007), we
obtained M(TRGB) = -4.11 at W606F -- W814F =1.00. Assuming E(B--V) = 0.032 from Schlafly \&
Finkbeiner (2011) for foreground reddening, we derived a distance
modulus of $(m-M)_0 = 29.05\pm0.11$, or the distance $D = 6.46\pm0.32$
Mpc. These values are somewhat larger than those obtained by Koda et al. (2015):
$(m-M)=28.61\pm0.23$ and $D=5.27\pm0.53$ Mpc from the photometry of the brightest 
stars resolved with the Subaru Prime Focus Camera.
The DOLPHOT photometry, full-field CMD, and a list of underlying
parameters are available on the Extragalactic Distance Database's
CMDs/TRGB Catalog (Anand et al. 2021).

\section{Ground based observations.}
  
 \subsection{Optical observations.}
 The galaxy KK~242 was observed on the 6-meter telescope of the
 Special Astrophysical Observatory (SAO) with the H$\alpha$ filter
 ($\lambda_{\rm eff}$ = 6555~\AA, $\Delta \lambda$ = 74~\AA) and in
 the mid-band filters SED 607 and SED 707 on April 22, 2018. The
 exposure time was 2$\times$600 sec in the H$\alpha$-line and
 2$\times$300 sec in the continuum. After standard data processing
 procedures, a faint emission spot with a flux of $F({\rm H}{\alpha})
 = (4.7\pm2.5)\times 10^{-16}$ erg/s/cm$^2$ was found on the western side of
 the galaxy (Kaisin \& Karachentsev 2019). This value is in
 satisfactory agreement with the H$\alpha$- flux estimate made by Koda
 et al. (2015) at the Subaru telescope.  The average of these two
 estimates corresponds to the star formation rate in KK~242 at a
 distance of 6.46 Mpc equal to SFR(H$\alpha$) = --4.82 dex
 ($M_{\odot}$/yr).  As noted by Koda et al. (2015), the galaxy was
 detected in the GALEX survey (Gil de Paz et al. 2007). With an
 apparent FUV-magnitude of (21.42$\pm$0.15) mag, its integral star
 formation rate is SFR(FUV) = --4.06 dex ($M_{\odot}$/yr).  The
 brightest part of the image of KK~242 in the FUV-band coincides with
 the HII-region visible in the H$\alpha$ line.
  
Recently, Pustilnik et al. (2021) performed spectral observations of
this HII-region with the SAO 6-m telescope and estimated the galaxy's
heliocentric radial velocity to be $V_h = -(65\pm25)$~km
s$^{-1}$. This optical velocity differs significantly from the old
velocity estimate of +426~km s$^{-1}$ obtained by Huchtmeier et
al. (2000) via the HI-line.

\subsection{The HI-observations with VLA.}

KK~242 was observed with the Karl G. Jansky Very Large
Array\footnote{The National Radio Astronomy Observatory is a facility
  of the National Science Foundation operated under cooperative
  agreement by Associated Universities, Inc.} (hereafter, VLA) in November 2019 for
program 19B-002 (P.I. Cannon).  Five hours of observing time were
acquired in three separate scheduling blocks.  The WIDAR correlator
was configured with a 16 MHz bandwidth centered on the expected
recessional velocity of KK~242 ($+$426 km\,s$^{-1}$; Huchtmeier et
al. 2000).  4096 spectral channels deliver a native spectral
resolution of 3.906 kHz\,ch$^{-1}$ (0.82 km\,s$^{-1}$\,cm$^{-1}$ at
the rest frequency of the 21\,cm line of neutral hydrogen).  The wide
bandwidth assures ample line-free channels for accurate continuum
subtraction.

Data reduction followed standard HI procedures.  Briefly, radio
frequency interference and bad data were excised.  The absolute flux
scale and bandpass shapes were determined by observations of the
standard calibrator 3C\,286.  Gain phases and amplitudes were
determined by observations of quasar J1748$+$7005.  The spiral galaxy
NGC\,6503 is within the VLA primary beam at 21\,cm when observing
J1748$+$7005.  The channels containing HI line emission from this
source (and the Milky Way) were blanked during the scans of the phase
calibrator field.  After calibration, continuum subtraction was
performed in the {\it uv} plane using a first-order fit to line-free
channels.

Imaging of the field was performed using the CASA TCLEAN algorithm
(see McMullin et al. 2007).  Calibrated and continuum-subtracted {\it
  uv} visibilities from all three datasets were imaged simultaneously
using the AUTO-MULTITHRESH algorithm (Kepley et al. 2020).  Initial HI
datacubes were centered at the expected HI velocity of KK~242 ($+$426
km\,s$^{-1}$; Huchtmeier et al. 2000).  Surprisingly, no HI emission
was detected within 200 km\,s$^{-1}$ at this location.  The final HI
cube had a synthesized beam size of
$\sim$60\arcsec$\times$$\sim$40\arcsec\
(asymmetric due to the northerly Declination of the source), a
velocity resolution of 5 km\,s$^{-1}$\,ch$^{-1}$, and reached an rms
noise of 9\,$\times$\,10$^{-4}$ Jy\,bm (within 7\% of the theoretical
noise level).  The initial conclusion was that the source was not
detected in the HI spectral line.

The recent downward revision of the recessional velocity of KK~242 by
Pustilnik et al. (2021) prompted us to re-examine the HI data products
with this prior.  As Figure 3 shows, the HI gas from NGC\,6503 and
from the Milky Way are both prominent in the velocity range between
$-$100 km\,s$^{-1}$ and 0 km\,s$^{-1}$.  These two contaminants have
high surface brightnesses over a range of angular scales, resulting in
a very complicated distribution of flux in the cubes.  Identifying the
HI flux from KK~242 is non-trivial.

After examining HI datacubes with a range of velocity resolutions, we
identified weak emission that is spatially coincident with the optical
position of KK~242 in the velocity range $-$90 km\,s$^{-1}$ to $-$70
km\,s$^{-1}$.  To maximize the S/N ratio of the HI gas from KK~242,
we created a datacube with 20 km\,s$^{-1}$ velocity resolution.  The
channel spacing was selected empirically to place all of the apparent
HI emission from KK~242 into a single channel of the resulting
datacube.  The AUTO-MULTITHRESH algorithm was used to identify signal
in the field (which can arise from the Milky Way, from NGC\,6503, or
from KK~242) which surpasses the 3.5\,$\sigma$ level in each channel.
Inside of these regions the cube was cleaned to a level of
0.5\,$\sigma$ (where $\sigma$ $=$ 4.5\,$\times$\,10$^{-4}$
Jy\,bm$^{-1}$ in 20 km\,s$^{-1}$ channels that are completely free of
HI emission).

We detect very faint HI emission that is spatially and spectrally
coincident with KK~242.  The source is visible in the center of the V
$=$ $-$80 km\,s$^{-1}$ panel of the channel maps shown in Figure 3.
Since the gridding of the HI data was chosen to encompass all of the
apparent HI emission from the source, this individual channel map can
be immediately converted into a moment zero image (representing total
HI mass surface density) by multiplying by the channel width.  Figure
4 shows this representation of the HI data, calibrated in units of HI
column density. The peak column density is a very modest
$\sim$3\,$\times$\,10$^{19}$ cm$^{-2}$.  The source is effectively
unresolved by the (asymmetric) beam.  Figure 5 shows the same HI
column density contours overlaid on a color image of KK~242 using
data from the Legacy Survey.  There is positional agreement of the HI
and optical components within half of the HI beam diameter.

The complicated nature of the HI in the spatial and spectral regions
surrounding KK~242 demands scrutiny about the reality of the source
and the significance of the detection.  To further examine the HI
properties, Figure 6 shows a position-velocity (PV) slice through the
HI datacube.  The $\pm$40\arcmin\ long PV slice was taken at an angle
of $+$270 degrees (measured eastward from north), was centered on the
HI maximum of KK~242 (determined from the moment zero map; see Figure
4), and was integrated over a width of 5 pixels (corresponding roughly
to the HI beam diameter).  The resulting PV slice shown in Figure 6
emphasizes that the HI gas that is associated with KK~242 is offset
both spectrally and spatially from the HI emission from the Milky Way
and NGC\,6503.

While Figures 3, 4, 5, and 6 are suggestive of the HI being associated
with the low-mass galaxy KK~242, the emission is of very low S/N
ratio and should be interpreted with caution.  Indeed, without the
velocity prior from Pustilnik et al. (2021), the putative source would
not have been identified.  Below we put forth some simplistic
interpretations of the HI properties on the assumption that the HI gas
is in fact associated with KK~242.  We stress that these
interpretations should be considered as demonstrative only.  More
sensitive observations are required to verify the putative detection.
The Huchtmeier et al. (2000) detection at +426 km s$^{-1}$ remains a puzzle
for us. If there was an HI signal of that strenght (2.03 Jy, km s$^{-1}$)
in the velocity range of the VLA then we would have detected it easily.

Integrating the HI emission in the single-channel moment map, we find
a total HI flux integral S$_{\rm HI}$ $<$ 0.1 Jy\,km\,s$^{-1}$ with an
error of no less than 50\%.  At the distance of 6.46 Mpc, the
corresponding HI mass is less than 10$^{6}$ M$_{\odot}$.  This HI mass
is comparable to that of the famous Local Volume galaxy Leo\,P
(McQuinn et al. 2015 and references therein) and lower than all of the
HI masses in the SHIELD program (McNichols et al. 2016; Teich et
al. 2016).  The S/N ratio of the HI data is too low to derive a
velocity field for KK~242, and so its rotational velocity and total
mass remain unconstrained by the HI data as presented here.

\section{Halo mass of NGC 6503 and other nearby low-mass spirals.}

Table 1 gives some basic properties of KK~242, including its
equatorial coordinates, morphological type, distance, heliocentric
radial velocity, apparent and absolute $B$-magnitudes, star formation
rates estimated via H$\alpha$ and FUV-fluxes, stellar mass, (putative)
HI-flux, and the total hydrogen mass.  Other photometric parameters of
the galaxy (its color, surface brightness, and effective radius) can
be found in Koda et al. (2015).

Assuming the Keplerian motions of small satellites around a central
massive galaxy with a typical eccentricity $e = 0.7$, we can estimate
the total mass of the main galaxy from the radial velocity difference,
$\Delta V$, and projected separation, $R_p$, of its satellite
(Karachentsev \& Kudrya 2014):

 $$ M_T = (16/\pi) G^{-1} \Delta V^2 R_p,$$ where $G$ is the
gravitation constant. This expression is based on the assumption that
the orbits of satellites are oriented randomly relative to the line of
sight.  For the pair of NGC~6503 and KK~242 at $\Delta V = 105$~km
s$^{-1}$ and $R_p = 31$ kpc, this mass estimate is $4.03\times10^{11}
M_{\odot}$. Obviously, this estimate is subject to significant
uncertainty due to the projection factor affecting $\Delta V$ and
$R_p$.  To get a more robust idea of the total mass of spiral galaxies
of moderate luminosity, we selected from the Updated Nearby Galaxy
Catalog (UNGC, Karachentsev et al. 2013) isolated Sc--Sd galaxies with
luminosities in the $K$-band $L_K/L_{\odot}$ = (9.5--10.0) dex, in
which small satellites have been found. In total, there are 9 such
galaxies in the Local Volume with a radius of 10 Mpc, around which 14
satellites are found, and 10 of them have measured radial
velocities. For all 9 spiral galaxies, the distances were measured
with an accuracy of $\sim$5\% by the TRGB method.  Data on these
galaxies and their companions are presented in Table 2. The Table
columns contain: (1) galaxy name; (2) morphological type according to
de Vaucouleurs scale; (3) galaxy distance; (4) method used to
determine the distance: via the tip of red giant branch (trgb), from
the Tully-Fisher relation (TF), based on a probable membership of the
dwarf galaxy in the suite of main galaxy (mem); (5) radial velocity of
galaxy relative to the Local Group centroid; (6) logarithm of integral
luminosity of galaxy in the $K$-band; (7) maximum amplitude of galaxy
rotation; (8,9) projected separation and radial velocity difference of
companion relative to the main galaxy; (10) estimate of the total
(orbital) mass. The data on the galaxies are taken from the last
version of the UNGC (http://www.sao.ru/lv/lvgdb) supplemented with
recent distance estimates from EDD. Several conclusions can be drawn
from the analysis of this data.
\begin{itemize}
  \item The root mean square radial velocity of satellites relative to
    the main galaxy is $\sigma_v = (63\pm7)$~km s$^{-1}$. Taking into
    account the average projection factor, $\sqrt{3}$, the
    characteristic spatial velocity of satellites, $(108\pm12)$~km
    s$^{-1}$, is similar to the average rotation amplitude of the
    spiral galaxies, $(94\pm8)$~km s$^{-1}$.

  \item The mean total (orbital) mass of the spiral galaxies of
    moderate luminosity is $\langle M_T/M_{\odot}\rangle =
    (3.46\pm0.84)\times 10^{11}$. The characteristic virial radius of the
    halo $R_{\rm vir}\simeq150$ kpc corresponds to this mass. The
    average projected separation of the satellites presented in Table
    2, $\langle R_p\rangle = (98\pm27)$ kpc, when multiplied 
by the projection factor ($4/\pi)$, is in good agreement
    with $R_{\rm vir}$.

  \item The mean luminosity of the considered spiral galaxies,
    $\langle L_K/L_{\odot}\rangle = (6.0\pm1.4)\times 10^9$, is about an
    order of magnitude lower than the luminosity of Milky Way and M
    31. The ratio of the average halo mass-to-average $K$-luminosity
    for them is $(58\pm19) M_{\odot}/L_{\odot}$, which is slightly
    higher than the $M_T/L_K$ for Milky Way,
    $(27\pm9)M_{\odot}/L_{\odot}$ and M\,31, $(33\pm6)
    M_{\odot}/L_{\odot}$(Karachentsev \& Kudrya 2014), also obtained
    from the motion of their satellites.
\end{itemize}

\section{Concluding remark.}
The isolated spiral galaxy NGC~6503 (a.k.a. KIG\,837 in the Catalog of
Isolated Galaxies; Karachentseva 1973), and its faint companion KK~242
are on the border of the Local Void (Tully 1988). According to Tully
et al. 2008, the center of the Local Void is at a distance of
$\sim$(10--20) Mpc from us, and the galaxies on the Void border are
moving from its center at a velocity of $\sim280$~km s$^{-1}$. Located
between the center of the expanding Void and the observer, the NGC~6503 \& KK~242 pair should have a negative peculiar velocity for
us. Indeed, at the measured distances and line-of-sight velocities and
adopting the Hubble parameter $H_0 = 73$~km s$^{-1}$Mpc$^{-1}$, the
peculiar velocities of both galaxies, $V_{\rm pec} = V_{LG} - H_0 D$,
are -173~km s$^{-1}$ and --294~km s$^{-1}$, respectively, confirming
the concept of the Local Void expansion.

Away from NGC~6503, opposite the Local Void, there is a chain of
groups around massive spiral galaxies: M~101, M~51 and M~63, the
distances $D$ and radial velocities $V_{LG}$ of which smoothly
increase with separation from NGC~6503 and the Local Void boundary:
6.95 Mpc and +375~km s$^{-1}$ (M 101), 8.40 Mpc and +553~km s$^{-1}$
(M 51) and 9.04 Mpc and 570~km s$^{-1}$ (M~63). This circumstance was
noted by M\"{u}ller et al. (2017). Interestingly, the distance (6.25
Mpc) and radial velocity (+283~km s$^{-1}$) of the NGC~6503 fits well
the configuration of this chain.  Together with NGC~6503, the angular
extent of this filament in the sky reaches about 30 degrees.

\acknowledgements 
{This work is based on observations made with the NASA/ESA Hubble Space
Telescope and with the Very Large Array.

J.M.C., J.F., and J.L.I. are supported by NSF/AST-2009894.

Support for program SNAP-15922 (PI Tully) was provided by NASA through
a grant from the Space Telescope Science Institute, which is operated
by the Associations of Universities for Research in Astronomy, 
Incorporated, under NASA contract NASb5-26555. I.D.K. is supported in part
by RNF grant 19--12--00145.}

\clearpage

\bigskip
{\bf  REFERENCES.}
\bigskip

Anand G.S., Rizzi L., Tully R.B., et al. 2021, arXiv:2104.02649

Bilicki M., Dvornik A., Hoekstra H., et al. 2021, arXiv:2101.06010

Carlin J.L., Sand D.J.,  Price P., et al. 2016, ApJ, 828L, 5

Carlsten S.G., Greco J.P., Beaton R.L., Greene J.E., 2020, ApJ, 891, 144

Correa C.A., Schaye J., 2020, MNRAS, 499, 3578

Crnojevich D., Sand D.J., Spekkens K., 2016, ApJ, 823, 19

Dolphin A.E., 2016, DOLPHOT: Stellar photometry, ascl:1608.013

Dolphin A.E., 2000, PASP, 112, 1383

Epinat B., Amram P., Marcelin M., et al. 2008, MNRAS, 388, 500

Gil de Paz A., Boissier S., Madore B.F., et al. 2007, ApJS, 173, 185

Greisen E.W., Spekkens K., van Moorsel G.A., 2009, AJ, 137, 4718

Huchtmeier W.K., Karachentsev I.D., Karachentseva V.E., Ehle M., 2000, A \& A Supplement, 141, 469

Kaisin S.S., Karachentsev I.D., 2019, Astr. Bull., 74, 1

Karachentsev I.D., Kudrya Y.N., 2014, AJ, 148, 50

Karachentsev I.D., Makarov D.I., Kaisina E.I., 2013, AJ, 145, 101

Karachentseva V.E., Karachentsev I.D., Melnyk O.V., 2011, AB, 66, 389

Karachentseva V.E., Karachentsev I.D., 1998, A \& A Supplement, 127, 409

Karachentseva V.E., 1973, Soobschenia Spec. Astrophys. Obs., 8, 3

Kepley, A.A., et al. 2020, PASP, 132, 1008, id 024505

Koda J., Yagi M., Komiyama Y., et al. 2015, ApJ, 801, L24

Kourkchi E., Tully R.B., 2017, ApJ, 843, 16

Lapi A., Salucci P., Danese L., 2018, ApJ, 859, 2

Makarov D.I., Makarova L.N., Rizzi L., et al. 2006, AJ, 132, 2729

Martin N.F., McConnachie A.W., Irwin M., et al, 2009, ApJ, 705, 758

Martinez Delgado D., Karim N., Boschin W., et al. 2021, arXiv:2104.03859

McMullin et al. 2007, Astronomical Society of the Pacific Conference
Series, 376, 127

McNichols et al. 2016, ApJ, 832, 89

McQuinn et al. 2015, ApJ, 812, 158

McQuinn et al. 2017, ApJ, 154, 51

M\"{u}ller O., Scalera R., Binggeli B., Jerjen H., 2017, A \& A, 602A, 119

Posti L., Fall S.M., 2021, A \& A, 649, A119 

Pustilnik S.A., Teplyakova A., Kotov S., 2021, MNRAS (in preparation)

Rizzi L., Tully R.B., Makarov D.I., et al. 2007, ApJ, 661, 815

Schlafly E.F., Finkbeiner D.P., 2011, ApJ, 737, 103

Teich et al. 2016, ApJ, 832, 85

Tollerud  E.J.,Beaton R.L., Geha M.C., et al, 2012, ApJ, 752, 45

Tully R.B., Shaya E.L., Karachentsev I.D., et al. 2008, ApJ, 676, 184

Tully R.B., 1988, Nearby Galaxy Catalog, (Cambridge: Cambridge Univ. Press) 

Williams B. F., et al., 2014, ApJS, 215, 9

Wu, P. F., Tully, R.B., Rizzi, L. et al. 2014, AJ, 148, 7

\clearpage

\begin{table}
\caption{Properties of KK~242.}
\begin{tabular}{lc}\\ \hline

 R.A. (J2000)         &           17 52 48.4
\\
 DEC (J2000)          &        +70 08 14
\\
 Type                 &                  Transition
\\
 $D$,   Mpc             &           6.46$\pm$0.32
\\
 $V_{\rm hel}$,  km s$^{-1}$         &          -80$\pm$10
\\
 $B$,  mag              &                18.6
\\
 $M_B$,    mag           &            -10.5
\\
 SFR(H$\alpha$), $M_{\odot}$/yr&   -4.82 dex
\\
 SFR(FUV),    $M_{\odot}$/yr &     -4.06 dex
\\
 $log(M_*)$, $M_{\odot}$            & 6.47 \\
S(HI),  Jy km s$^{-1}$          &  $<0.1 $ \\
$log(M_{HI})$,    $M_{\odot}$   &  $< 6.0 $ \\
\hline
\end{tabular}
\end{table}

\begin{table}
\begin{small}
\caption{Detached spiral Sc--Sd type galaxies in the Local Volume with
 $\log(L_K/L_{\odot}) = 9.5-10.0$, and their probable companions.}
\begin{tabular}{lrclrrrrrr}\\ \hline
 Name          &    T &   $D$&    meth &$ V_{LG}$& $\log L_K$&  $V_{\rm rot}$ & $R_p$&  $\Delta V$&    $M_T$
 \\
\hline

               &       & Mpc  &        & km s$^{-1}$ & $L_{\odot}$&   km s$^{-1}$ & kpc&  km s$^{-1}$ &$10^{10} M_{\odot}$
\\
\hline 
(1)&(2)&(3)&(4)&(5)&(6)&(7)&(8)&(9)&(10)\\
   \hline
 M 33            &  6  &  0.93 &  trgb &   34 &   9.62  &   99 &     0  &   0  &   \\
 And XXII        & -3  &  0.79 &  trgb &   87 &   5.28  &      &    47  &  53  &  15.6\\
 Tri III         & -3  &  0.82 &  trgb &    - &   4.43  &      &    82  &   -  &    - \\   
 NGC 2403        &  6  &  3.19 &  trgb &  262 &   9.86  &  128 &     0  &   0  &     \\
 MADCASHJ0742+65 & -2  &  3.39 &  trgb &   -  &   5.86  &      &    36  &   -  &   -\\
 DDO 44          & -3  &  3.21 &  trgb &  356 &   7.78  &      &    73  &  94  &  76.1\\
 NGC 2366        &  9  &  3.28 &  trgb &  251 &   8.70  &      &   206  & -11  &   2.9\\
 NGC 7793        &  6  &  3.63 &  trgb &  250 &   9.70  &  101 &     0  &   0  &   \\
 PGC 704814      & 10  &  3.66 &  trgb &  299 &   6.90  &      &    14  &  49  &   4.0\\
 NGC 1313        &  7  &  4.31 &  trgb &  264 &   9.57  &  120 &     0  &   0  &   \\
$[$KK98$]$27         &  10 &   4.23&   trgb&   327&    7.04 &      &     25 &   63 &   11.7 \\
 NGC 4236        &   8 &   4.41&   trgb&   157&    9.61 &    70&      0 &    0 &  \\
 $[$KK98$]$125       &  10 &   4.40&   mem &    - &    6.61 &      &     50 &    - &    -\\
 DDO 165         &   9 &   4.83&   trgb&   196&    8.23 &      &    374 &   39 &   67.1\\
 NGC 5068        &   6 &   5.15&   trgb&   469&    9.73 &    66&      0 &    0 &      \\
 dw1318-21       & -1  &  5.15 &  mem  &   -  &   7.06  &      &    78  &   -  &   - \\
 NGC 6503        &   6 &   6.25&   trgb&   283&   10.00 &    91&      0 &    0 &  \\
 $[$KK98$]$242       & 10  &  6.46 &  trgb &  178 &   6.47  &      &    31  & -105 &  40.3\\
 NGC 4656        &  8  &  7.98 &  trgb &  635 &   9.93  &   64 &     0  &   0  &    \\
 NGC 4656UV      & 10  &  7.98 &  mem  &  568 &   8.88  &      &    40  & -67  &  21.2\\
 NGC 7640        &  6  &  8.43 &  trgb &  668 &   9.77  &  107 &     0  &   0  &   \\ 
 UGC 12588       &  8  &  8.43 &  mem  &  723 &   8.83  &      &   103  &  55  &  36.8\\
 DDO 217         &  8  &  8.55 &  TF   &  720 &   9.37  &      &   219  &  52  &  69.9\\
\hline
Average &-&-&-&-&-& 94$\pm$8& 98$\pm27$& +22$\pm$21& 34.6$\pm$8.4\\ 
\hline

\end{tabular}
\end{small}
\end{table}
\clearpage

 \begin{figure}
   \begin{center}\includegraphics[width=14cm]{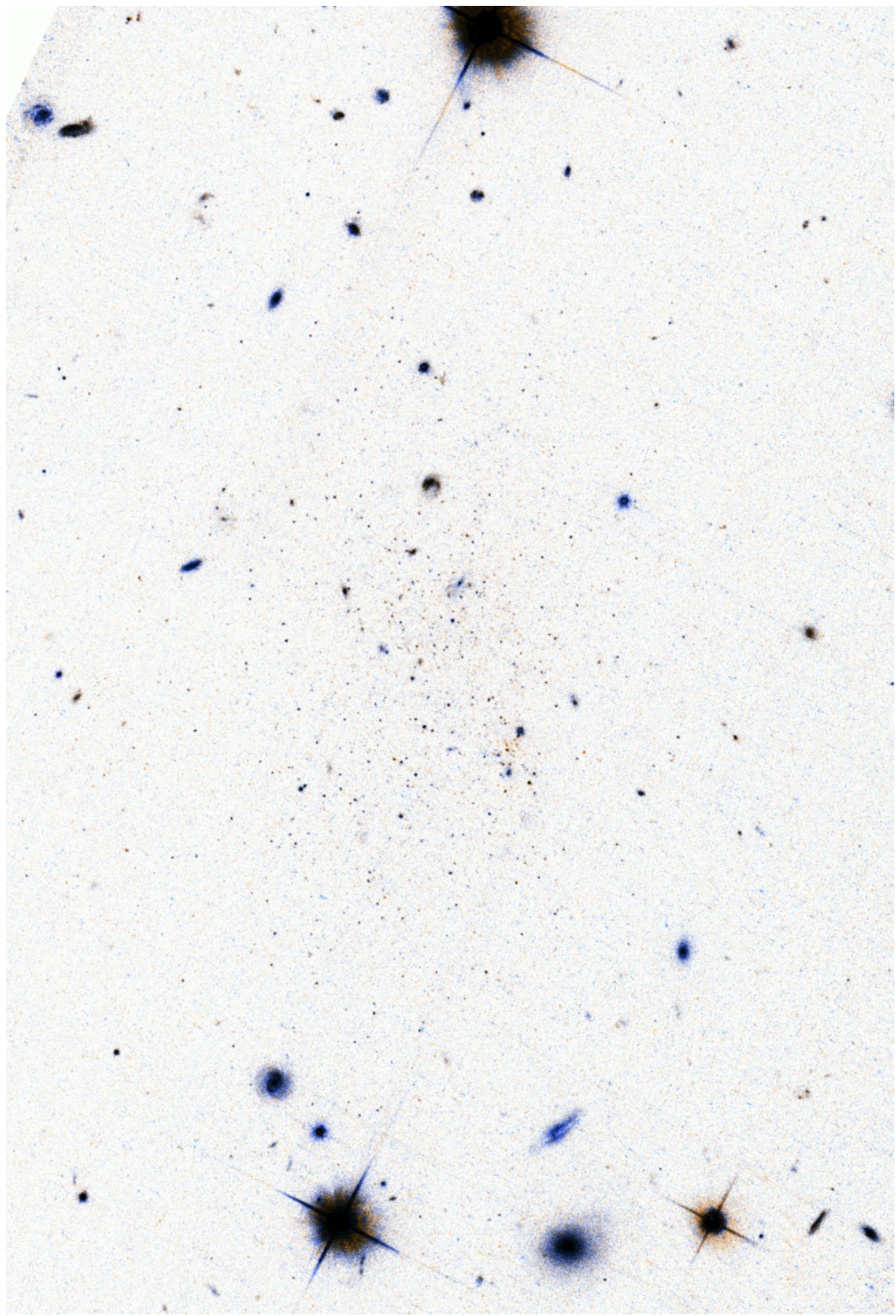}\end{center}
    \caption{HST/ACS combined image of KK~242. The image size is
      104$\times70\arcsec$. North is up and east is left.  The
      inverted color image is composite of images with F606W and
      F814W filters. Blue stars are seen as red dots.}
 \end{figure}
 \clearpage

\begin{figure}
  \includegraphics[width=17cm]{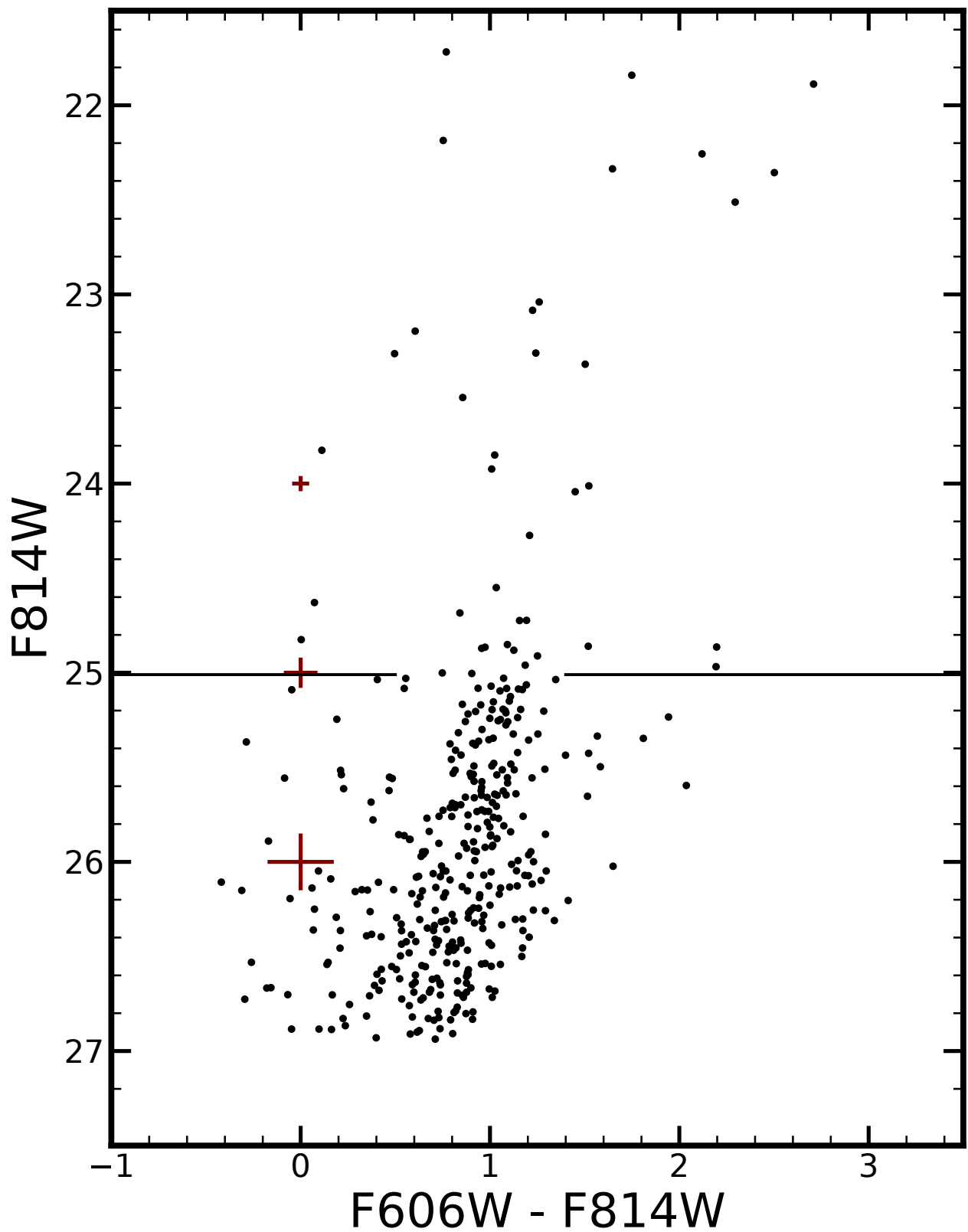}
    \caption{Color-magnitude diagram of KK~242. The TRGB position is
      indicated by the horizontal black line. Representative error
      bars (calculated at the observed color of the TRGB) are shown in
      maroon.}
\end{figure}
\clearpage

\begin{figure}
  \includegraphics[width=17cm]{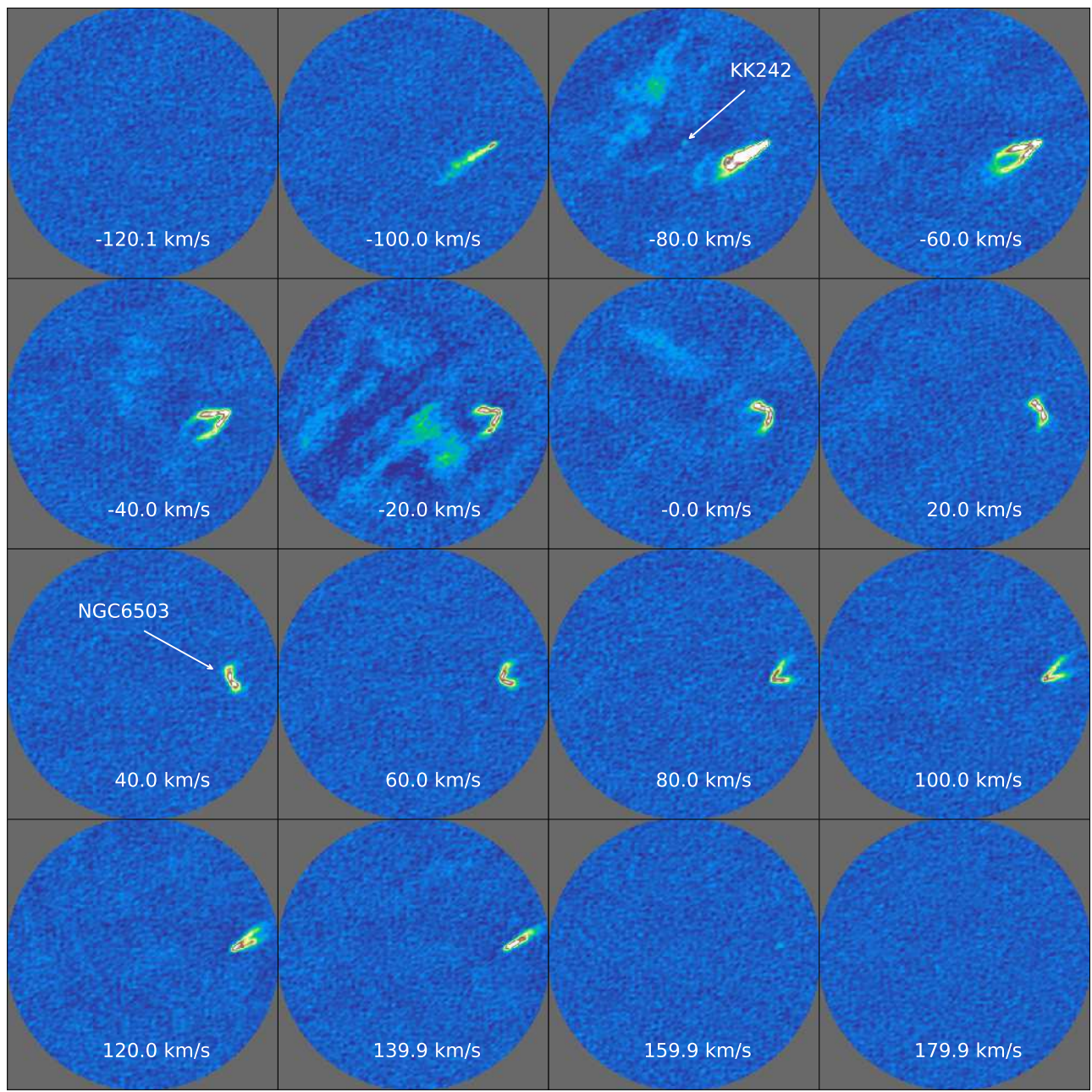}
 \caption{Channel maps showing HI emission from KK~242, the Milky
   Way, and NGC\,6503.  The velocity resolution is 20
   km\,s$^{-1}$\,ch$^{-1}$, and the intensity scale spans the range
   from 50\% of the (negative) minimum value in the cube to 50\% of
   the (positive) maximum value in the cube.  Valid pixels are shown
   within the VLA primary beam to the level of 5\% of the maximum.
   Strong HI emission is detected from the field spiral NGC\,6503;
   weaker but more widespread HI emission from the Milky Way is
   prominent.  The tentative detection of HI gas from KK~242 is
   apparent in the $-$80 km\,s$^{-1}$ panel; the compact HI source at
   the field center is co-spatial with the optical body and
   co-spectral (wtihin errors) with the emission line velocity derived
   by Pustilnik et al. (2021). The field of view is 53.3\arcmin on a
   side.}
\end{figure}
\clearpage

\begin{figure}
  \includegraphics[width=17cm]{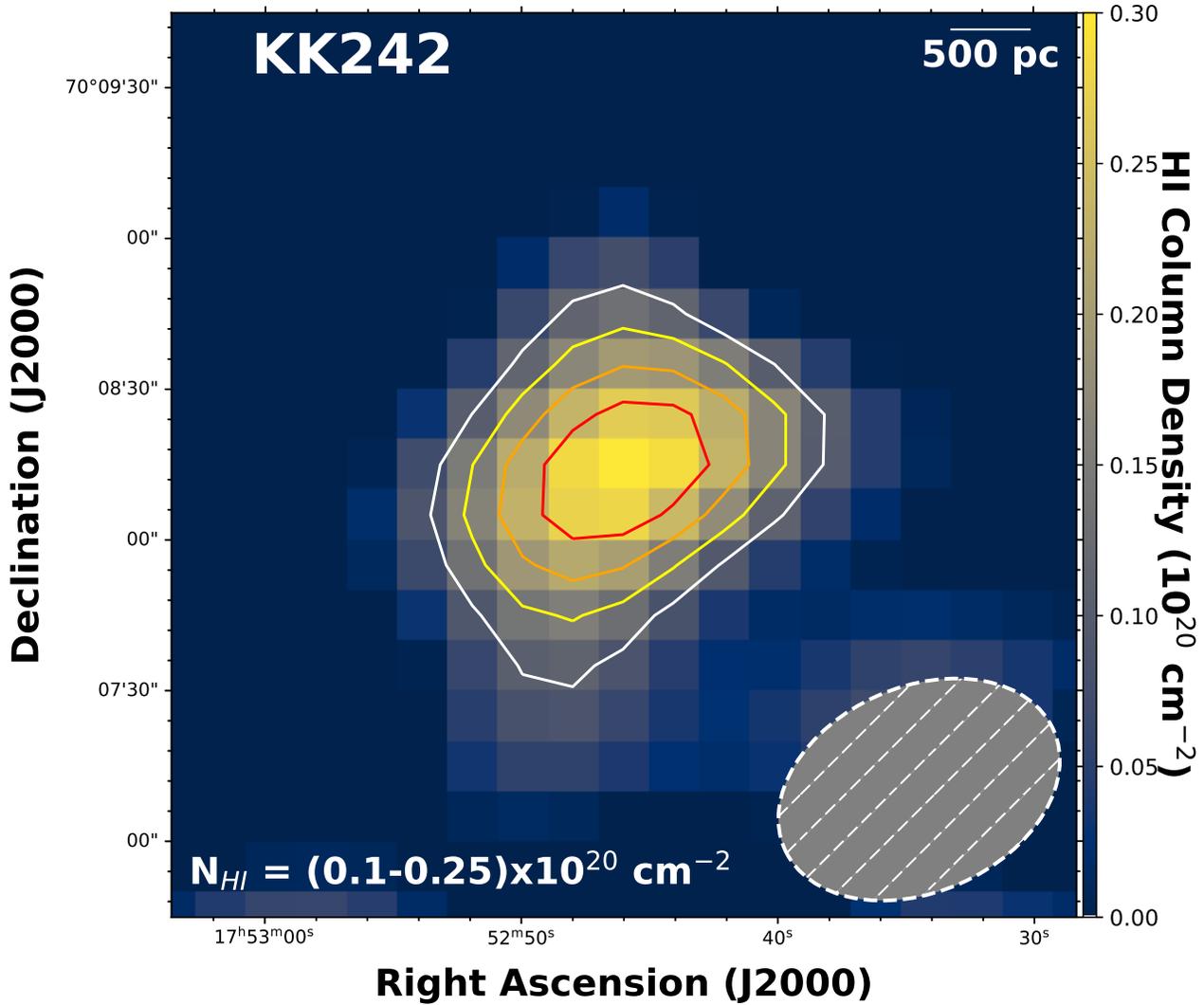}
  \caption{Integrated HI image of KK~242, created by imaging the HI
    in the velocity range $-$90 km\,s$^{-1}$ to $-$70 km\,s$^{-1}$ and
    enforcing a 3.5\,$\sigma$ threshold.  The putative HI emission is
    very faint and of low S/N.  Contours are shown in column density
    units at levels of (0.1, 0.2, 0.3)\,$\times$\,10$^{20}$ cm$^{-2}$.
    The synthesized beam size of 58.5\arcsec\ $\times$ 40.8\arcsec\ is
    shown by the shaded ellipse at the lower right.}
\end{figure}
\clearpage

\begin{figure}
  \includegraphics[width=17cm]{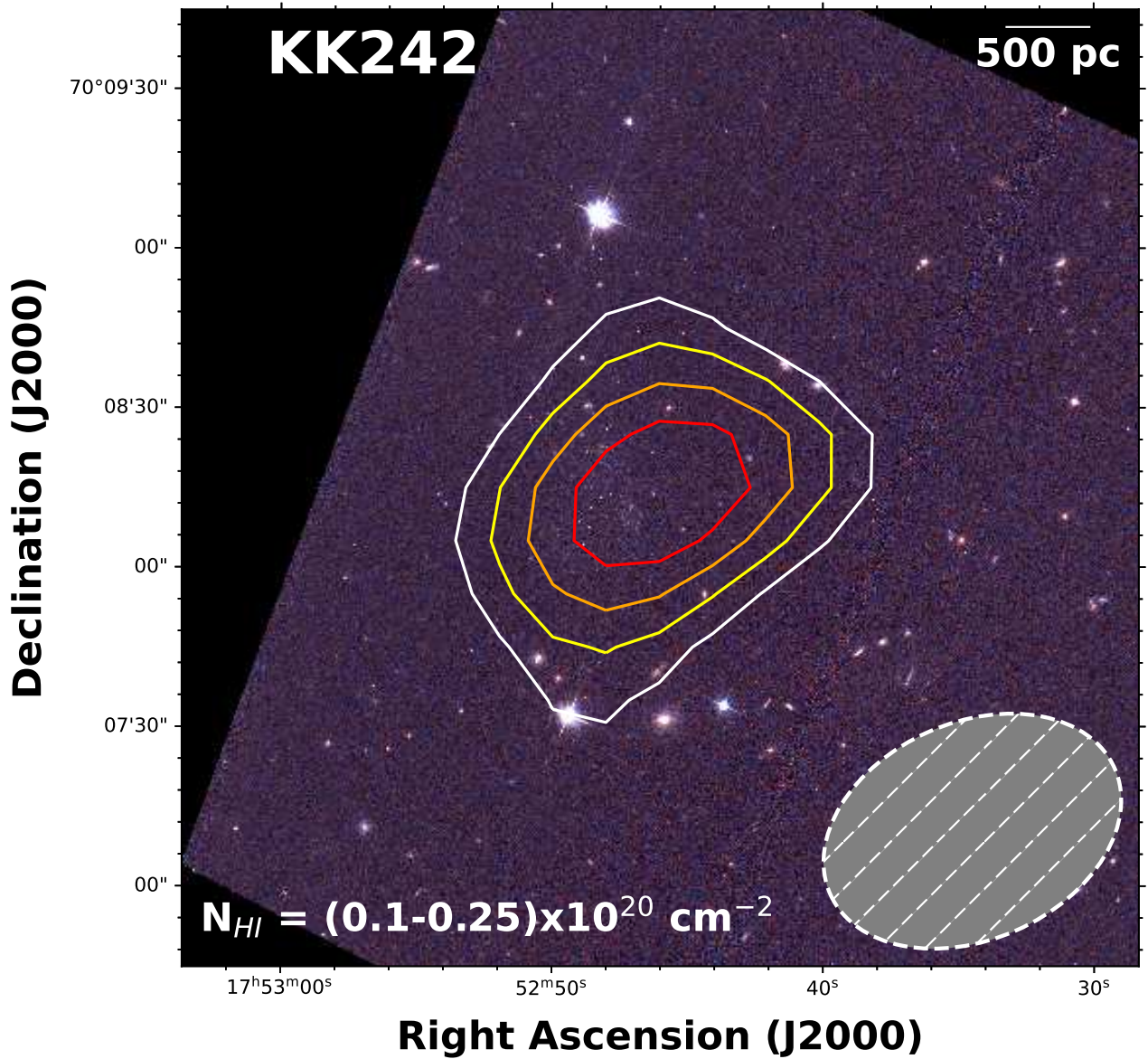}
 \caption{HI column density contours (from Figure 4) overlaid on an
   HST three-color image of KK~242.  The HI centroid is co-spatial
   with the optical counterpart within one half of the HI beam.  The
   synthesized beam size of 58.5\arcsec\ $\times$ 40.8\arcsec\ is
   shown by the shaded ellipse at the lower right.}
\end{figure}
\clearpage

\begin{figure}
  \includegraphics[width=18cm]{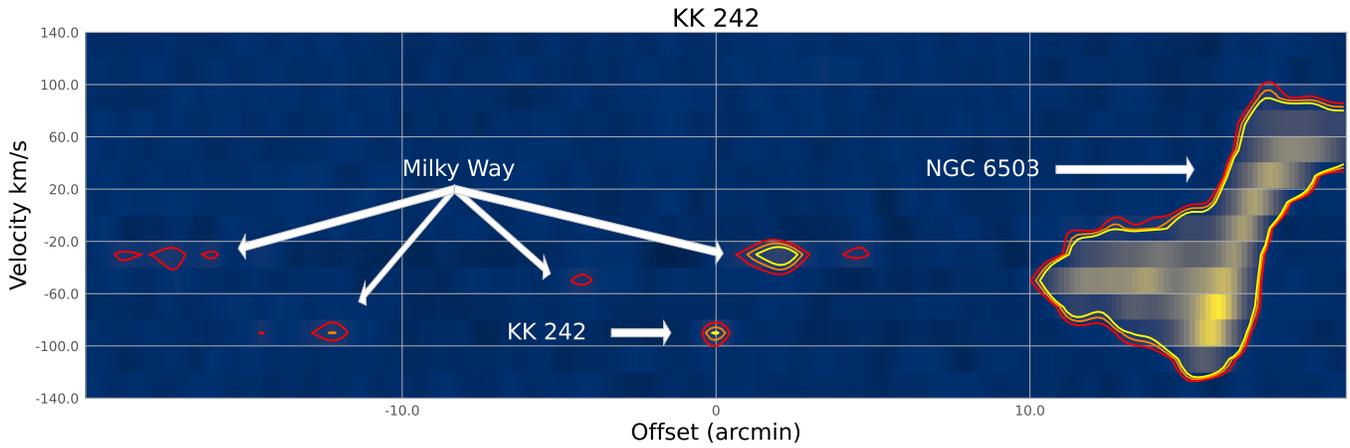}
 \caption{P-V slice through the HI datacube shown in Figure 3.  The
   50\arcsec\ wide slice is oriented East-West, passes through the HI
   centroid of KK~242 (see Figure 4), and extends $\pm20\arcmin$.
   The slice intersects the putative HI gas associated with KK~242 as
   well as with the disk of NGC\,6503 and with multiple foreground
   Milky Way HI clouds.  Contours are shown at levels of 4, 5.5, and 7
   $\sigma$.  This P-V diagram suggests that the faint HI emission
   that is co-spatial with the optical body of KK~242 is distinct in
   both position and velocity space from other HI structures in the
   field.}
\end{figure} 

\end{document}